# High resolution 3D laser scanner measurements of a strike-slip fault quantify its morphological anisotropy at all scales

François Renard

Laboratoire de Géophysique Interne et Tectonophysique, Université de Grenoble, France & Physics of Geological Processes, University of Oslo, Norway

Christophe Voisin

Laboratoire de Géophysique Interne et Tectonophysique, Université de Grenoble, France

David Marsan

Laboratoire de Géophysique Interne et Tectonophysique, Université de Savoie, France

Jean Schmittbuhl

Institut de Physique du Globe de Strasbourg, UMR 7516, Strasbourg, France

**Abstract.** The surface roughness of a recently exhumed strike-slip fault plane has been measured by three independent 3D portable laser scanners. Digital elevation models of several fault surface areas, from 1 m$^2$ to 600 m$^2$, have been measured at a resolution ranging from 5 mm to 80 mm. Out of plane height fluctuations are described by non-Gaussian distribution with exponential long range tails. Statistical scaling analyses show that the striated fault surface exhibits self-affine scaling invariance with a small but significant directional morphological anisotropy that can be described by two scaling roughness exponents, $H_1 = 0.7$ in the direction of slip and $H_2 = 0.8$ perpendicular to the direction of slip.

## 1. Introduction

Fault morphology is often oversimplified at small scales in most fault models or earthquake simulations despite its influence on the mechanical behavior of the fault. Indeed asperities on active faults planes concentrate the stress and therefore control earthquake nucleation [Scholtz 2002]. Moreover striations on a fault plane can show substantial rotation during single earthquake because of changes of slip with time during the rupture propagation see Spudich et al. [1998], and references therein]. Unfortunately, fault morphology is rather complex over a large range of scales [Power et al, 1987] but of small magnitude which makes the imaging of the fault plane very difficult at depth. Only high resolution relocations of earthquakes using the multiplet technique have shown earthquake alignments along faults. This pattern was interpreted as resulting from the presence of an organized roughness (asperities) resisting the slip [Schaff et al. 2002].

A quasi-unique access to high resolution description of the fault plane comes from fault scarp observations. Owing to technical limitations, the roughness of several fault planes has only been studied using 1D profilometry. From these pioneer measurements, fault roughness has been shown to be scale invariant with roughness exponent close to 0.8 [Power et al.

1987, 1991; Schmittbuhl et al. 1993; Power and Durham, 1997], a property similar to fresh mode I fracture surfaces [Schmittbuhl et al, 1995]. Surprisingly, the scaling invariance appears to be close to isotropic [Power et al, 1988; Lee and Bruhn 1996] though it has been shown that normal fault surfaces contain striations developing parallel to the direction of slip and holes elongated in a direction perpendicular to the slip. Such features might be scale dependent.

Among the limitations of previous studies, measurements were performed mainly on faults with small cumulated slip. It is not common to have access to well preserved exhumed fault planes with a large cumulated slip. For instance, normal faults provide large scarps because of the vertical motion of the fault and if the motion is fast enough, scarps might be suited for roughness measurements. The total slip is however limited. Larger accumulated slips can be achieved on strike-slip faults but, as the vertical displacement is small, active fault scarps can seldom be observed. In this study, one such fault plane has been measured. The Vuache fault, near Annecy in the French Alps, is 30 km long and has accumulated a couple of kilometers of left-lateral slip. The fault zone contains several parallel fault planes and the measured plane has therefore accumulated only an undetermined fraction of the total displacement. Some minor right-lateral slips are also witnessed by shear criteria on outcropping fault planes. It is still an active fault: it broke in a $M_L$=5.3 earthquake located at 2 km depth on 07/16/1996, 16 km from the outcrop studied [Thouvenot et al. 1998]. The second important aspect of this study is that the outcrop of this fault has been sampled not only along 1D profiles but using three 3D laser scanners (also called LIDAR, Light Detection And Ranging) which provide new high resolution statistics of the fault roughness.

## 2. 3D scanner data of the Vuache fault

The outcrop studied (GPS coordinates N 45°57.242', E 6°02.933') consists of a 60x20 m plane cutting Cretaceous limestones of decimeter scale bedding. It was selected because it was exhumed 10 years ago by the activity of a quarry, and was thus almost perfectly preserved from atmospheric alteration or glacial erosion. Striations at all scales on the fault plane, dipping at 15-30°, indicate a strike-slip motion, with a small normal component. The striations are parallel to the bedding and the largest grooves are larger than the sedimentary beds. We assume here that bedding does not the scaling relationships of the roughness described below.

The fault surface was measured using three independent portable 3D laser scanners (Table 1). One device (S10) scans the surface by triangulation using a red laser and a camera. The other two devices (GS100, LMS Z420i) measure the time of flight of an infra-red reflected laser spot. Six sub-surfaces of the fault were scanned at various resolutions to cover surface scales from ~1 $m^2$ to 600 $m^2$ at a resolution from 5 mm to 60 mm, coarser than the precision of the scanner.

The result is a cloud of points whose three-dimensional coordinates correspond to the points on the fault surface with a regular angular spacing (Figure 1a). This 3D set of points is transformed into a 2+1D data set corresponding to a Digital Elevation Model (DEM) over the mean plane of the fault (Figure 1, bottom). From the DEM data sets, 1D profiles can be extracted or the whole 2D surface can be directly analyzed using various statistical tools detailed below.

## 3. 1D surface roughness analysis

A set of parallel cuts is taken through the digitized surface to obtain a series of parallel profiles striking at an angle $\theta$ from the horizontal. Linear detrending is performed independently on all the profiles to yield a set of heights $z_i$, function of the coordinate $x_i$ along the cut. The statistical properties of these 1D profiles are then calculated and averaged over a large number of parallel profiles. The manner in which the scaling varies with $\theta$, and hence with the orientation relative to the direction of slip, was also studied.

### 3.1. Height distribution

The distribution of height difference $\delta z = |z_i - z_j|$, for two intervals of $\delta x = |x_i - x_j|$ is shown on Figure 2. At small height difference ($|\delta z|$ < 20 mm) the distribution is approximately Gaussian, which is likely to be due to the noise in the data. The distribution however significantly departs from a Gaussian law, admitting large values much more frequently than a Gaussian model would predict, with a decay for large values of $\delta z$ that roughly follows an exponential distribution. This feature is seen at all values of $\theta$. The existence of exponentially distributed large height differences covering about 1% of the fault surface area implies that purely Gaussian models (e.g. fractional Brownian motions) of the fault roughness would fail to capture the largest geometrical asperities.

### 3.2. 1D analysis: root-mean-square correlation functions

The roughness of the fault is studied by looking at how the distributions of height difference vary with spatial wavelength. The aim is to demonstrate that the surface remains unchanged under the scaling transformation $\delta x \to \lambda \delta x$, $\delta z \to \lambda^H \delta z$ for 1D profiles extracted from the surface [Meakin, 1998]. Here, $\delta x$ is the coordinate along the profile and $\delta z$ is the roughness. For a self-affine profile, the Hurst exponent $H$ lies in the range $0 \leq H \leq 1$. Moreover, as the surface is striated along a well-defined direction, anisotropic scaling behavior can emerge if $H$ varies relative to the direction of slip.

A set of parallel profiles, with a constant orientation relative to the striation was extracted. For all the profiles at a given angle relative to the direction of slip, the standard deviation $\sigma(\delta x)$ of the height $\delta z$ was computed for different intervals of $\delta x$. Figures 3a&b show the growth of the roughness $\sigma(\delta x)$ for slip-parallel and slip-perpendicular profiles. The analysis was performed on three scans, acquired with the three different scanners (Table 1). Levelling off of $\sigma(\delta x)$ at small $\delta x$ values is due to the noise. For a Gaussian white noise, with standard deviation $\sigma_{noise}$, the measured $\sigma(\delta x)$ verifies:

$$\sigma^2(\delta x) = \sigma_0^2(\delta x) + 2\sigma_{noise}^2. \quad (1)$$

This formula can be used to remove the influence of the noise to obtain the corrected standard deviation $\sigma_0(\delta x)$. Self-affine profiles are defined over three decades by a single scaling behavior $\sigma_0(\delta x) \propto \delta x^H$ where $H$ is the Hurst exponent (Figure 3).

Focusing on a single scan, the data show that there is a length scale close to 1 cm where the roughness does not significantly depend on the orientation relative to the slip and the surface is isotropic (Figure 4a).

A systematic increase in the Hurst exponent is observed when the angle of the profile departs from the slip-parallel orientation (Figure 4b). Parallel to the striation, the Hurst exponent is close to 0.8, and lower than the value close to 0.9 found for profiles oriented perpendicular to the slip.

## 4. 2D height-height difference analysis

A more thorough 2D analysis of the anisotropy involves computing the best fitting planes $y = a_i x + b_i z + c_i$ centered on all points $(x_j, y_j, z_j)$ of the fault, using a least-mean square method with a weight on point $(x_j, y_j, z_j)$ following an exponential function $exp(-r_{ij}/L)$ where $r_{ij}$ is the distance between the two points and $L$ is the scale of observation. It is then possible to estimate how the height root-mean-square changes with distance, along any direction $\theta$, without resorting to 1D profiles. This is achieved by computing the root-mean-square (rms) of the height difference $(a_i \cos\theta + b_i \sin\theta)/L$. This has been applied on a striated surface (Fig. 5a) where this rms is shown for varying angles and scales (Fig. 5b). The changes in roughness with $\theta$ is clear, with the smoothest fault geometry being obtained at an angle of nearly 20°. This confirms the presence of anisotropy at all scales, down to the resolution of the scanner. There is no characteristic wavelength for the striation.

## 5. Discussion & conclusion

According to Figures 3 and 4, the 1D roughness can be described as $\sigma(\delta x) = \alpha(\delta x)^H$. Both the pre-factor $\alpha$ and the Hurst exponent $H$ depend on the angle $\theta$, with the smallest value of $H$ in the direction of slip (Figure 4b). The angular variation of $H$ has also been observed at a smaller scale on experimental shear fractures [Amitrano and Schmittbuhl, 2002], where $H$ was equal to 0.74 in the slip direction and 0.8 perpendicular to this direction. This is consistent with observations reported by *Power and Durham* [1997] and *Lee and Bruhn* [1996], and confirmed by the 3D data here.

In addition, the two power laws of figure 4a cross at a specific length scale close between 1cm and 1 mm. Below this length scale, the fault surface is rougher in the direction of slip, and above this length scale, the fault is rougher perpendicular to the direction of slip. However, there is no direct link between this length scale and the characteristic weakening distance observed or estimated for the nucleation process.

Although the direction of slip clearly appears on the 3D data, the statistical tools described here are not sufficient to determine whether the slip was left-lateral or right-lateral. This opens up a future challenge to develop statistical tools that can target the asymmetry of the slip and, therefore, the geometry of the asperities resisting the slip.

**Acknowledgments.** This project has been supported by the CNRS (ACI and DyETI programs) and the University J. Fourier (BQR). We would like to thank J.-M. Nicole for his technical help, and F. Thouvenot, J.-P. Gratier, M. Bouchon, and G. Puaux for helpful discussions.

## References

Amitrano, D., and J. Schmittbuhl, Fracture roughness and gouge distribution of a granite shear band, J. Geophys. Res., 107, 2375, 2002.

Lee, J.-J., and R. Bruhn, Structural anisotropy of normal fault surfaces, J. Struc. Geol., 18, 1043-1059, 1996.

Meakin, P., Fractals: scaling and growth far from equilibrium, Cambridge Univ. Press, New York, 1998.

Power, W. L., T. E. Tullis, S. R. Brown, G. N. Boitnott and C. H. Scholz, Roughness of natural fault surfaces, Geophys. Res. Lett., 14, 29-32, 1987.

Power, W. L., T. E. Tullis, and J.D. Weeks, Roughness and wear during brittle faulting, J. Geophys. Res., 93, 15268-15278, 1988.

Power, W. L., and T. E. Tullis, Euclidean and fractal models for the description of rock surface roughness, J. Geophys. Res., 96, 415-424, 1991.

Power, W. L., and W. B. Durham, Topography of natural and artificial fractures in granitic rocks: Implications for studies of rock friction and fluid migration, Int. J. Rock Mech. Min., 34, 979-989, 1997.

Schaff, D., G. H. R. Bokelmann, G. C. Beroza, F. Waldhauser, and W. L. Ellsworth, High resolution image of Calaveras fault seismicity, J. Geophys. Res., 107, 633-668, 2002.

Schmittbuhl, J., S. Gentier, and S. Roux, Field measurements of the roughness of fault surfaces, Geophys. Res. Lett., 20, 639-641, 1993.

Scholz, C.H. The mechanics of earthquake and faulting, Cambridge University Press, Campbridge, 2002.

Spudich, P. Guatteri, P., Otsuki, K., and J. Minagawa, Use of fault striations and dislocations models to infer tectonic shear stress during the 1995 Hyogo-ken Nanbu (Kobe) earthquake, Bull. Seismol. Soc. Am., 88, 413-427, 1998.

Thouvenot, F., J. Fréchet, P. Tapponnier, J. C. Thomas, B. Le Brun,, G. Ménard, R. Lacassin, L. Jenatton, J.R. Grasso, O. Coutant, A. Paul, and D. Hatzfeld, The $M_L$ 5.3 Epagny (French Alps) earthquake of 1996 July 15: a long-awaited event on the Vuache Fault, Geophys. J. Int., 135, 876-892, 1998.

David Marsan, Laboratoire de Géophysique Interne et Tectonophysique, Université de Savoie, 73376 Le Bourget du Lac Cedex, France

François Renard & Christophe Voisin, Laboratoire de Géophysique Interne et Tectonophysique, BP 53, 38041 Grenoble, France

Jean Schmittbuhl, Institut de Physique du Globe de Strasbourg, 5 rue René Descartes, 67084 Strasbourg cedex, France

**Figures & Table**

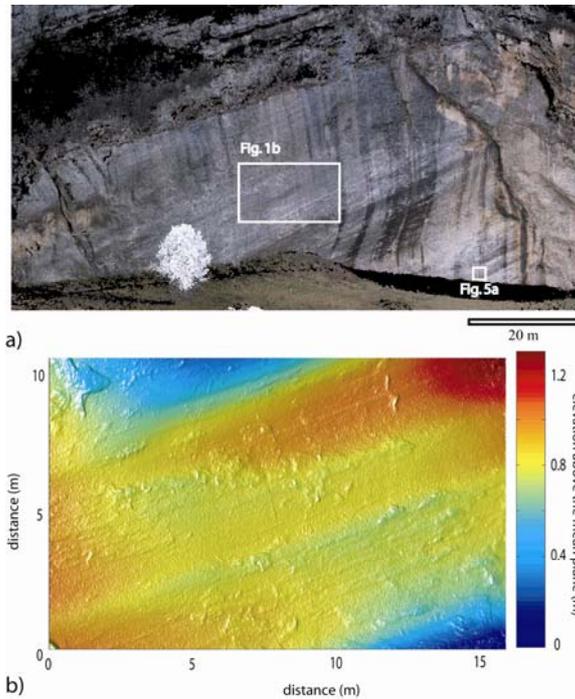

**Figure 1.** Scanner 3D data of the Vuache exhumed fault surface. a) Whole outcrop view. The insets correspond to the surfaces shown on Fig. 1b and 5a. b) Zoom on the fault. The striation, with maximum amplitude of around 1.2 m, can be detected at all scales up to the measurement resolution. The surface contains 434,600 points, on a constant grid of 20x20 mm (measurements done on a 10x10 mm grid and then averaged on a coarser grid). The resolution of the elevation measurement is 4.5 mm.

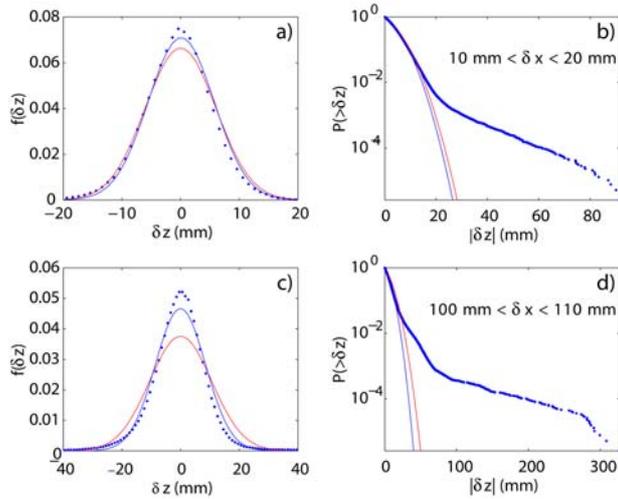

**Figure 2.** Left: Distribution of the height difference $\delta z$ for two intervals (a) $\delta x$=10-20 mm and (c) $\delta x$=100-110 mm, for $\theta$=30° and data shown on Fig. 1b. Right (b-d): Logarithm of the probability of exceedence, roughly showing the exponential decay of the distribution for the last ~1% of the largest slopes. Blue dots: data. Red lines: best Gaussian fit for the whole distribution. Blue lines: best Gaussian fit for $\delta z$ within 3 standard deviations of its mean. Most of the heights $\delta z$ appear to follow the reduced Gaussian fit (blue), but with clear departure at large absolute values $|\delta z|$.

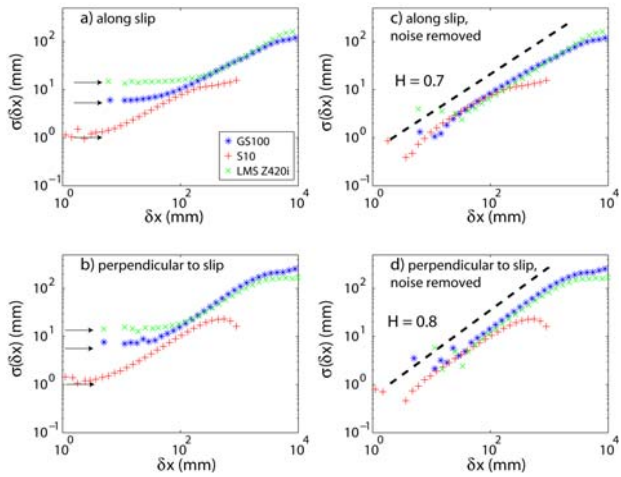

**Figure 3.** (a, b) rms of the height difference $\sigma(\delta x)$ along two directions, parallel and perpendicular to the direction of slip. (c, d) rms of $\sigma_0(\delta x)$, the height difference after removing the noise, as defined by Eq. (1). The three data sets correspond to the three scanners used in this study (Table 1). The level of noise for each scanner is estimated as the height of the flat part of the height distribution at small $\delta x$ values and is indicated by the arrows. As $\delta x \to 0$, $\sigma(\delta x) \to \sqrt{2}\,\sigma_{noise}$. The noise rms $\sigma_{noise}$ is found to be independent of the angle $\theta$ of the cut, as expected for an isotropic white noise. The flattening of the scaling behavior at large scales is related to a finite size effect. Data of the surface shown on Figures 1b & 5a.

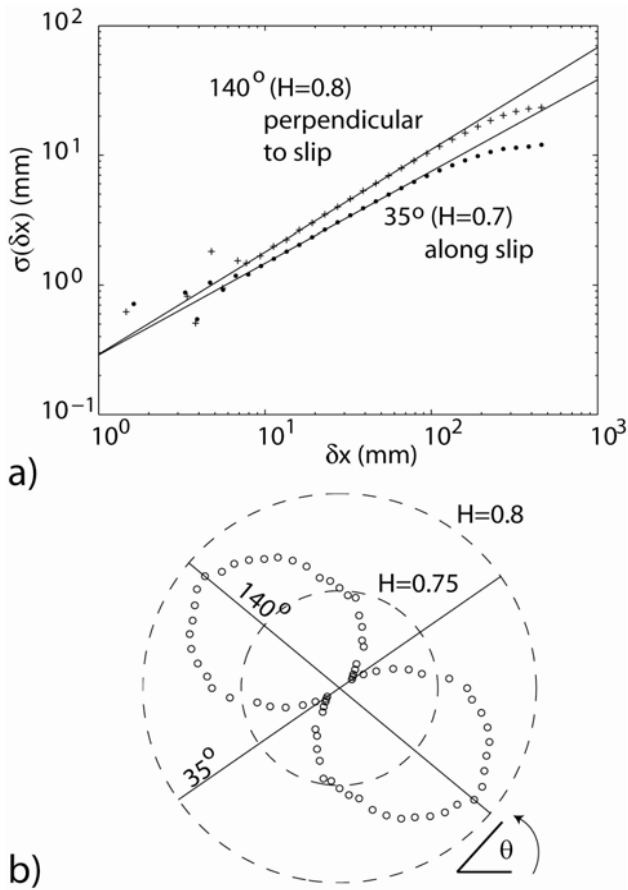

**Figure 4.** Surface anisotropy revealed by the angular variation of the Hurst exponent. a) Root-mean-square of the height-height correlation as a function of the distance along two profiles in the direction of the striation (35°) and roughly perpendicular to it (140°), showing the variation of the Hurst exponent. b) Polar plot of H. The minimum Hurst exponent (H=0.7) is at 35°, in the direction of the slip, and the maximum (H = 0.8) is in an almost perpendicular direction. Data of the surface shown on Figure 5a.

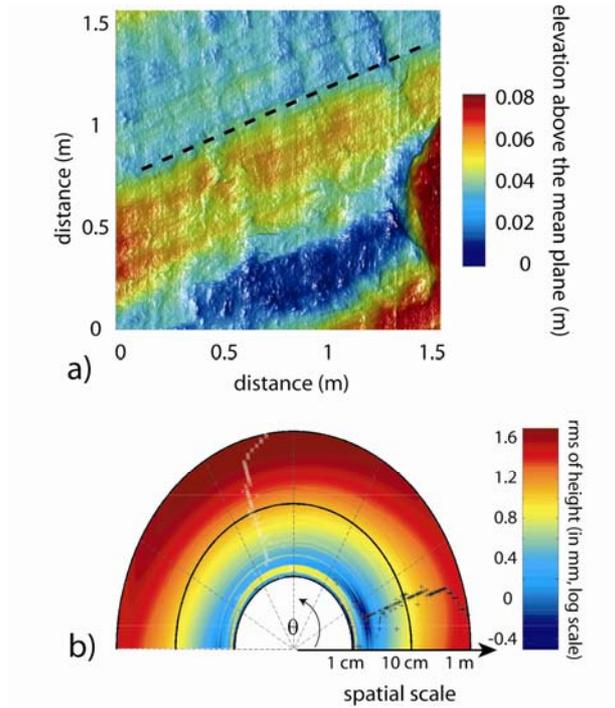

**Figure 5.** Surface anisotropy revealed by 2D analysis. a) DEM of the surface at a resolution of 5 mm. The dashed lines give the angle of the striation relative to the horizontal. The surface contains 320,250 points, on a constant grid of 5x5 mmm. The resolution of the elevation measurement is 0.9 mm. b) Root-mean-square of the heights $\delta z$ in mm resulting from a 2D analysis (see text). The influence of noise is removed according to equation (1). The minima (black +) and maxima (white x) of the root-mean-square are tracked with the scale. The minimum roughness is observed along slip, at about 20° from the horizontal.

**Table 1.** Laser scanner characteristics.

| 3D scanner device | S10 | GS100 | LMS Z420i |
|---|---|---|---|
| **Company** | ©TRIMBLE | ©TRIMBLE | ©RIEGL |
| **Resolution ($\delta x$)** | 5 mm | 10 and 20 mm | 60 mm |
| **Noise on the data** | 0.9 mm | 4.5 mm | 10.2 mm |
| **Acquisition speed** | 70 pts/s | 5000 pts/s | 5000 pts/s |
| **Data recovery** | 99%[1] | > 99.5% | > 99.5% |

[1] The measurements were performed during the night to avoid interference with sun light. The level of noise is calculated on the data (see section 3.2)